\documentstyle[preprint,revtex]{aps}
\begin{document}
\draft
\begin{title}
An Electrostatic Model of Split-Gate Quantum Wires
\end{title}
\author{Yinlong Sun and George Kirczenow}
\begin{instit}
Department of Physics, Simon Fraser University\\
Burnaby, British Columbia, Canada V5A 1S6  
\end{instit}
\author{Andrew. S. Sachrajda and Yan Feng}
\begin{instit}
Institute of Microstructural Sciences, National Research Council \\ 
Ottawa, Ontario, Canada  K1A 0R6 
\end{instit} 
\receipt{December 12, 1994}
\begin{abstract}
We present a theoretical model of split-gate quantum wires that are 
fabricated from GaAs-AlGaAs heterostructures. 
The model is built on the physical properties of donors and of 
semiconductor surfaces, and considerations of equilibrium in such systems.
Based on the features of this model, we have studied different 
ionization regimes of quantum wires, provided a method to evaluate 
the shallow donor density, and calculated the depletion and pinchoff 
voltages of quantum wires both before and after illumination.
A real split-gate quantum wire has been taken as an example for 
the calculations, and the results calculated for it agree well with 
experimental measurements.
This paper provides an analytic approach for obtaining much useful 
information about quantum wires, as well as a general theoretical 
tool for other gated nanostructure systems.
\end{abstract}
\pacs{PACS number: 73. 20. Dx}

\section{Introduction}
\label{sec:introduction}
Modern material-growing techniques such as molecular beam epitaxy 
and organo-metallic chemical vapour deposition make it possible 
to fabricate extremely clean semiconductor heterostructures.\ \cite{Chang}
In a modulation-doped \cite{Dingle} GaAs-${\rm Al_{x}Ga_{1-x}As}$
 heterostructure, a two-dimensional electron gas (2DEG) is present 
at the interface of ${\rm Al_{x}Ga_{1-x}As}$ and GaAs 
layers.\ \cite{Harris}
This 2DEG can be further confined laterally by various confining 
techniques such as electron-beam lithography \cite{Wheeler}, 
ion-beam exposure \cite{Scherer,Hirayama,Wieck}, or 
etching \cite{Skocpol,Kirtley,van Houten}, forming a 
quasi-one-dimensional system usually called a quantum wire.
At present, one widely used confinement method is the 
split-gate technique \cite{Thornton,Zhang}.
In a split-gate quantum wire, when a sufficiently negative 
voltage is applied to the metallic gates, electrons are 
completely depleted from under the gates, leaving a central 
channel of electrons undepleted.
By further increasing the gate voltage negatively, the density 
of electrons in the 
channel is decreased continuously until the channel pinches off.
Such quantum wires display unique fascinating properties which 
have stimulated many theoretical and experimental studies of 
their physics.\ \cite{Ulloa}
Because of the sophisticated gating technique and the 
flexibility of changing the density of electrons by varying 
the gate voltage, the split-gate quantum wires hold a great 
potential for realistic applications \cite{Sheard,Khuraba}.
 
Considerable progress has been made in developing an understanding 
of the electronic structure of quantum wires theoretically, based 
on the results of computer simulations and analytic 
work.\ \cite{Lai,Laux86,Laux88,Davies,Nixon,Nakamura,Ravaioli,Sun93}
However, for many systems, particularly those with an exposed 
semiconductor surface between the split metallic gates, the 
current understanding is not complete.
For example, it has not been possible to predict accurately 
the pinchoff voltage of a quantum wire, given the knowledge of 
the geometric and doping parameters and the history of a given sample.
The pinchoff voltage is the gate voltage at which conduction 
through the wire ceases.
It is a quantity of considerable practical importance for 
these devices.

In this paper, we present a study that addresses such issues.
In Section~\ref{sec:model description} we describe the basic physical
features of gated quantum wires that are included in our model. 
In Sections~\ref{sec:regimes} and \ref{sec:determination} we point
out that three qualitatively different ionization regimes can exist in
the doped layer that supplies electrons to the quantum wire, and show
how the ionization regime that a particular sample is in can be
identified. We also show how the shallow donor density in the doped
layer may be calculated.  
In Section~\ref{sec:depletion} we describe the
calculation within our model of the depletion voltage for the electron
gas under the gates. 
In Section~\ref{sec:pinchoff} we calculate the
pinchoff voltage of the quantum wire.  
This calculation uses a Green's
function method, which is an extension of the previous theoretical work
of Davies \cite{Davies} but treats the effects of the charges at
the exposed semiconductor surface more accurately. 
In Section~\ref{sec:discussion}, we take a well-characterized real
sample as an example for calculations, and find good
quantitative agreement between the calculated and measured depletion
and pinchoff voltages both before and after illumination.

\section{Model and Formalism}
\label{sec:model}
We consider an infinitely long split-gate quantum wire whose 
crossection is shown in FIG.~\ref{fig:crossection}.
The layers from top are the GaAs cap, the Si-doped ${\rm Al_{x}Ga_{1-x}As}$, 
the undoped ${\rm Al_{x}Ga_{1-x}As}$ spacer, and the GaAs channel; 
their thicknesses are $l_{c}$,  $l_{d}$, $l_{s}$, and $l_{ch}$, 
respectively.
On top of the GaAs cap are two metallic gates with a spatial 
separation $2w$.
The coordinate frame is chosen in such a way that the exposed 
surface of GaAs cap is the $z=0$ plane, and the lateral direction 
is along the x-axis.

In such quantum wires, electrons donated by the Si donors in the 
doped ${\rm Al_{x}Ga_{1-x}As}$ layer transfer to the $z=0$ plane 
to fill the surface or interface states, and to the 
$z=L=l_{c}+l_{d}+l_{s}$ plane to form the 2DEG. 
This transfer of electrons leaves a positive spatial charge in 
the doped layer and thus causes the conduction band to bend within 
the heterostructure.
One possible case of the band bending is shown in FIG.~\ref{fig:regime C}.
The curve depicts to the bottom of the conduction band along the z-axis.
Within the cap and the spacer layers, the curve is linear 
because there is no spatial charge in these layers.				
In the doped layer, however, the bottom of the conduction band 
is curved because of the presence of the spatial charge.
The curve is parabolic if the spatial charge density is uniform.
$E_{off}$ is the conduction band offset which occurs at the two 
interfaces between the GaAs and ${\rm Al_{x}Ga_{1-x}As}$ layers.
The whole system shown is in equilibrium, that is, the system has 
a uniform Fermi energy.
(This may change when a voltage is applied between the gates and 
2DEG, as is discussed below.)
Note that, in the situation shown in FIG.~\ref{fig:regime C}, 
there is an unionized region in the doped layer where the 
conduction band is flat because donors are not ionized there.
A detailed discussion of the features of our model now follows.

\subsection{Model Description}
\label{sec:model description}
Our theoretical model of split-gate quantum wires has four key features.

{\bf Feature 1) 
The Si donors are uniformly distributed in the ${\bf Al_{x}Ga_{1-x}As}$ 
doped layer and divided into two categories: the shallow levels and 
the deep levels.
We assume that electrons are donated only by the ionized shallow 
donors whose bound energy levels are above the Fermi level.
Deep donors can be ionized by illumination.}

It is well-known that the electronic state associated with a 
shallow donor in ${\rm Al_{x}Ga_{1-x}As}$ has the hydrogenic 
form and can be handled with the effective mass theory \cite{Kohn}.
Neglecting central cell effects, the binding energy of a shallow 
donor is 
\( E_{s} = m^{*} e^{4}/2(4\pi \varepsilon \varepsilon_{0} 
\hbar )^{2}= m^{*} / \varepsilon^{2} {\rm (Ryd)} \), 
where $m^{*}$ is the effective mass of the electron and $\varepsilon$ 
is the dielectric constant.
In ${\rm Al_{x}Ga_{1-x}As}$, the $\Gamma$ valley of the conduction
 band is the lowest one when $x<0.45$.
In quantum wires, $x$ is usually in this regime.
At the mininum point of the $\Gamma$ valley, 
$m^{*}=0.067\ m_{e}$ and $E_{s} \approx 6$ meV correspondingly.
Such a binding energy of shallow donors has been verified by 
various measurements \cite{Lifshitz,Ishikawa}. 
Because $E_{s}$ is much less than other relevant parameters 
such as the Schottky barrier and the conduction band offset, 
we consider $E_{s}$ to be negligible small.

In the doped ${\rm Al_{x}Ga_{1-x}As}$ layer, when $x>0.2$, 
the ground state of a Si donor is the deep level instead of 
the shallow level.\ \cite{Chand}
It is now generally accepted that the deep level is associated 
with a local lattice distortion which is usually called a 
DX center \cite{Lang92,Malloy}.
During illumination, a deep donor may absorb a photon and thus ionize.
At low temperatures, however, a shallow donor can not change 
into a deep donor automatically because of the energy barrier 
associated with the lattice distortion.
This argument is supported by many studies such as 
persistent-photoconductivity experiments \cite{Lang77}.

Accordingly, we have
\begin{equation}
N_{total} = N_{s} +N_{d},
\label{donor sum}
\end{equation}
where $N_{total}$, $N_{s}$, and $N_{d}$ are the total, shallow, 
and deep donor concentrations, respectively.
For a quantum wire, $N_{total}$ can be obtained from the fabrication 
parameters but $N_{s}$ and $N_{d}$ are undetermined experimentally.
This has made it difficult to analyze quantum wires theoretically 
because $N_{s}$ determines the number of donated electrons and the 
spatial charge density.
However, we will describe a method to calculate $N_{s}$ within 
our model.

{\bf Feature 2) 
The Schottky barrier between the metallic gates and the GaAs cap 
is determined by the type of the gate metal and the type of GaAs 
interface, and is independent of the gate voltage.
The surface states of the exposed GaAs surface are pinned at a 
single energy level within the forbidden band gap of GaAs.
The surface states are localized and surface electrons have a 
low mobility.}

The Schottky barrier of a metal-semiconductor contact refers to 
the energy difference between the conduction band minima of the 
semiconductor at the interface and the Fermi level of electrons 
in the metal. 
It is generally believed \cite{Monch} that Schottky barriers are 
associated with the metal-induced gap states which depend only on 
the type of the contact metal and the type of the semiconductor 
interface.
This means that, in quantum wires, the Schottky barrier between 
the gates and the GaAs cap is independent of the gate voltage.
For (100) and (110) interfaces of GaAs, the Schottky barriers for 
many metals have been measured \cite{Waldrop,Newman,McLean}.
 
The surface states of GaAs are associated with the dangling bonds 
at the exposed surface.
The physics of surface states is complicated and there has been no 
generally accepted model yet.\ \cite{Monch}
However, experiments show that the surface states of GaAs are pinned 
at a single energy value within the forbidden band gap as long 
as the surface is covered by a fraction of an adatom 
monolayer.\ \cite{Spicer}
For example, the surface states of the n-type GaAs (100) surface 
are pinned at about 0.8 eV below the bulk conduction band 
minima.\ \cite{Chiang}
Some calculations \cite{Potz,Beres} also show that the surface 
states are very localized, which means that the surface 
electrons have a very low mobility.

The Schottky barrier of the exposed surface refers to the energy 
difference between the conduction band minima and the pinned 
surface level (see FIG.~\ref{fig:regime C}).
In the following discussion, we use $\Phi_{sb}$ for the Schottky 
barrier of the exposed surface and $\Phi_{sb}'$ for the Schottky 
barrier of the metal-GaAs contact.

{\bf Feature 3)
The energy barrier due to the spacer layer is small. 
Therefore we assume that the electrons on either sides of it are 
always in equilibrium with each other. 
The energy barrier that separates the surface electrons is so high 
that tunneling of electrons through it can be neglected.
Therefore we assume that the total number of surface electrons 
is conserved when the gate voltage varies.}

This feature can be justified by that the tunneling 
current of
electrons through an energy barrier is proportional to the tunneling
probability of an electron through the barrier.
In the WKB approximation, the tunneling probability of an electron 
at the Fermi level is 
\begin{equation}
T=\exp [-2 \int_{\rm (barrier)} dz \sqrt{2m^{*}(E_{c}(z)-E_{F}) 
\over \hbar^{2}}],
\end{equation}
where $E_{c}(z)$ is the conduction band minimum 
(refer to FIG.~\ref{fig:regime C}).
Because the energy barrier due to the spacer in a typical quantum wire is
small ($l_{s}=20$ nm and $E_{off}=0.2$ eV typically), the corresponding
tunneling current is so large that it keeps electrons on both sides
in equilibrium no matter how the gate voltage changes. 
On the other hand, the barrier at the exposed surface is very
high ($l_{c}=10$ nm, $l_{d}=40$ nm, and $\Phi_{sb}=0.8$ eV typically),
therefore the corresponding tunneling current is so small that the
surface electrons are isolated and the total number of the surface
electrons is conserved although the gate voltage changes.

{\bf Feature 4)
We assume that, after a quantum wire has been fabricated and no 
gate voltage is applied, the surface electrons share the same 
Fermi energy with the 2DEG.
We also assume that this equilibrium also holds after the 
quantum wire undergoes illumination at the zero gate voltage.}

This assumption is based on the consideration that the high-temperature 
($T\sim$ 500 K) fabrication process provides the conditions necessary 
for the whole system to reach equilibrium.
That is, the surface electrons share the same Fermi energy 
with the rest of the system.
After the quantum wire is illuminated, the surface electrons 
do not necessarily stay in equilibrium with the others.
However, by assuming the equilibrium of the whole system 
after illumination, we have a starting point for calculation 
of the effect that an illumination has on the quantum wires.
Moreover, we speculate that the real situation of 
quantum wires after an illumination by photons with energies 
larger than $\Phi_{sb}$ is not too far from an 
equilibrium state, and therefore the evaluated results 
should provide useful information.

Based on the four features presented above, we are able 
to set up the electrostatic formalism for any quantum wire 
system and make predictions.
However, we need first to determine the shallow donor 
density $N_{s}$, which is not directly known from the 
sample fabrication conditions or from experimental measurements.
We find that $N_{s}$ can be determined from $n_{0}$, 
the 2DEG density at the zero gate voltage, which can be 
obtained by extrapolating the densities measured from 
edge state backscattering experiments~\cite{Haug,Washburn} 
to zero gate voltage.
However, the relation between $N_{s}$ and $n_{0}$ depends 
on the ionization regime of shallow donors in the doped 
${\rm Al_{x}Ga_{1-x}As}$ layer.
Therefore, we need to analyze the ionization regimes 
of the doped layer at zero gate voltage.
 
\subsection{Ionization Regimes}
\label{sec:regimes}
The ionization regimes here refer to the spatial arrangement 
of the ionized shallow donors in the doped layer, and to the 
way that the donated electrons are distributed between the 2DEG 
and the surface states, at zero gate voltage.
For the quantum wire shown in FIG.~\ref{fig:crossection}, there 
are three ionization regimes.

In ionization regime A, the band bending is shown in 
FIG.~\ref{fig:regimes AB}a.
In this regime, all of the shallow donors in the doped layer are 
ionized and no 2DEG is present.
Because the bottom of the conduction band in the GaAs channel 
layer is higher than the surface (interface) levels, all donated 
electrons transfer to the $z=0$ plane to fill the surface (interface) 
states.
The electrons accumulated at the $z=0$ plane in effect form 
a `capacitor' with the positively ionized donors in the doped layer.
Thus the conduction band in the GaAs channel layer is not affected 
by the transfer of electrons and remains flat.
The ionization of the doped layer falls into this regime when the 
shallow donor density is very low.

In ionization regime B, the band bending is shown in 
FIG.~\ref{fig:regimes AB}b.
In this regime, all the shallow donors in the doped layer are 
ionized and a 2DEG is formed at the $z=L$ plane.
Note that the curved conduction band within the doped layer 
has a minimum point M which divides the whole doped layer 
into two parts, with thicknesses $l_{1}$ and $l_{2}$, respectively.
Because the electric field at M is zero, one may consider 
all of the donated electrons from the region to the left of
M to transfer to the $z=0$ plane thus form a `capacitor', 
while all the donated electrons to the right of M transfer to 
the $z=L$ plane to form another `capacitor'.
These two capacitors have no interaction each other because 
each screens itself completely.
Such a consideration enables us to discuss each capacitor separately.

Ionization regime C is the most complicated and its band bending 
structure is shown in FIG.~\ref{fig:regime C}.
Regime C differs from regime B by the presence of an 
{\em unionized region} in the doped layer.
In the unionized region, the bound levels of shallow donors are not 
above the Fermi level, thus the electrons in this region are not 
ionized and remain bound to the donors.
Correspondingly, the whole doped layer is divided into three parts.
The left hand one forms one `capacitor' with the surface 
(interface) electrons, the right hand one forms another 
`capacitor' with the 2DEG, and the central one is charge neutral 
with its conduction band being flat.
The ionization occurs in regime C when the shallow donor density 
is very high.

For the quantum wire shown in FIG.~\ref{fig:crossection} with fixed 
geometric dimensions, as the shallow density increases, the ionization 
of the doped layer progresses from regime A to B, to C.
In studying quantum wires, however, we are only interested in 
the ionization regimes B and C when the 2DEG is present. 
Usually the ionization is in regime B before illumination 
and in regime C after sufficient illumination, because the shallow 
donor density is increased by illumination.

To identify the ionization regime of a particular quantum wire 
at the zero gate voltage, we need to calculate the critical 
characteristic parameters.
Notice that since the exposed surface Schottky barrier $\Phi_{sb}$ and 
the metal-GaAs contact Schottky barrier $\Phi_{sb}'$ may be quite 
different, we have to calculate the critical parameters under 
the gates and under the exposed surface separately.
For under the exposed surface, let $N_{\alpha}$ be the critical 
shallow donor density that divides regimes A and B, and $N_{\beta}$ be 
the one that divides regimes B and C.
Under the gates, let $N_{\alpha}'$ and $N_{\beta}'$ be the  
corresponding critical parameters.

Now let us calculate $N_{\alpha}$ and $N_{\beta}$.
When $N_{s} = N_{\alpha}$, the bottom of conduction band in the 
GaAs channel layer (the region denoted `flat' in 
FIG.~\ref{fig:regimes AB}a) lines up with the system's Fermi level, 
and with the energy level of the surface states.
Therefore,
\begin{equation}
{e^2 N_{\alpha} \over \varepsilon \varepsilon_{0} }
( l_{c}l_{d} + { l_{d}^2 \over 2} ) = \Phi_{sb},
\label{eq:Nalpha}
\end{equation}
in which the left side gives the total band bending in the cap 
and the doped layers.
The two band offsets at $z=l_{c}$ and $z=L$ cancel each other.
(Here, as well as in following discussion, we take the 
`capacitor' to be large and neglect its edge effects.)

When $N_{s}=N_{\beta}$, the minimum point M in 
FIG.~\ref{fig:regimes AB}b just touches the x-axis.
That is, the bottom of the conduction band at M is at the 
system's Fermi level.
Therefore we have
\begin{eqnarray}
{e^2 N_{\beta} \over \varepsilon \varepsilon_{0} }
( l_{c}l_{1} + { l_{1}^2 \over 2} ) & = & \Phi_{sb} + E_{off}, 
\label{eq:Nbeta-1} \\
{e^2 N_{\beta} \over \varepsilon \varepsilon_{0} }
( l_{s}l_{2} + { l_{2}^2 \over 2} ) & = & E_{off} - E_{z0}, 
\label{eq:Nbeta-2} \\
l_{1} + l_{2} & = & l_{d},
\label{eq:Nbeta-3}
\end{eqnarray}
where equations \ref{eq:Nbeta-1} and \ref{eq:Nbeta-2} come from 
the fact that the bottom of conduction band at point M is equal to 
the Fermi energy of surface electrons and of the 2DEG.

Note that, in equation \ref{eq:Nbeta-2}, $E_{z0}$ is the energy 
difference between the 2DEG Fermi energy and the bottom of 
the conduction band at $z=L$, as shown in FIG.~\ref{fig:regimes AB}b.
Typically $E_{z0}\sim 0.04$ eV but $E_{z0}$ vanishes when electrons 
are nearly depleted.\ \cite{Harris}
Because $E_{z0}$ is comparable to $E_{off}$, we include its effect 
in equation~\ref{eq:Nbeta-2}.
($E_{z0}$ does not appear in equation~\ref{eq:Nalpha}, because electrons 
are depleted in that situation.)

When $N_{s}=N_{\beta}$, the corresponding density of surface electrons is
\begin{equation}
n_{\beta}^{sur} = N_{\beta} l_{1},
\label{eq:nsurbeta}
\end{equation}
and the density of the 2DEG is
\begin{equation}
n_{\beta} = N_{\beta} l_{2},
\label{eq:nbeta}
\end{equation}
where $N_{\beta}$, $l_{1}$, and $l_{2}$ are obtained by solving equations 
\ref{eq:Nbeta-1}, \ref{eq:Nbeta-2}, and \ref{eq:Nbeta-3}.

For the doped layer under the gates, the critical parameter values 
$N_{\alpha}'$ and $N_{\beta}'$ can be calculated similarly except 
that the surface states are replaced by the interface states and 
$\Phi_{sb}$ by $\Phi_{sb}'$.

We can identify the ionization regime of a particular quantum wire by 
comparing its actual shallow donor density $N_{s}$ to its calculated 
critical values $N_{\alpha}$ and $N_{\beta}$.
However, $N_{s}$, being a part of $N_{total}$, is usually not known directly.
On the other hand, the 2DEG density $n_{0}$ at zero gate voltage can 
readily be determined experimentally.
Therefore, it is more convenient to work in terms of the comparison 
between $n_{\beta}$ and $n_{0}$.
The conditions for different ionization regimes of the doped layer 
{\em under the exposed surface} are listed in
TABLE~\ref{tab:ionizations}. 
(The conditions for the ionization regimes
of the doped layer {\em under the gates} are obtained by replacing
$N_{\alpha}$, $N_{\beta}$, and $n_{\beta}$ in 
TABLE~\ref{tab:ionizations} by primed
quantities.)

\subsection{Determination of $N_{s}$}
\label{sec:determination}
Now we evaluate $N_{s}$ from the measured 2DEG density $n_{0}$  at 
zero gate voltage.
For a quantum wire in ionization regime B, $N_{s}$ is obtained by 
solving the following equations
\begin{eqnarray}
{e^2 N_{s} \over \varepsilon \varepsilon_{0} }
( l_{c}l_{1} + { l_{1}^2 \over 2} ) 
- {e^2 N_{s} \over \varepsilon \varepsilon_{0} }
( l_{s}l_{2} + { l_{2}^2 \over 2} ) 
& = & \Phi_{sb} + E_{z0}, 
\label{eq:rhos-1} \\
l_{1} + l_{2} & = & l_{d}, 
\label{eq:rhos-2} \\
N_{s} l_{2} & = & n_{0},
\label{eq:rhos-3}
\end{eqnarray}
where $l_{1}$ and $l_{2}$ have been shown in FIG.~\ref{fig:regimes AB}b.
Equation~\ref{eq:rhos-1} comes from the condition that the surface 
energy level is equal to that of the 2DEG.

If the ionization of the quantum wire is in regime C 
(see FIG.~\ref{fig:regime C}), then $N_{s}$ 
should be calculated from
\begin{eqnarray}
{e^2 N_{s} \over \varepsilon \varepsilon_{0} }
[l_{s}l_{2} + { l_{2}^2 \over 2} ] 
& = & E_{off} - E_{z0}, 
\label{eq:rhos-4} \\
N_{s} l_{2} & = & n_{0},
\label{eq:rhos-5}
\end{eqnarray}
where equation~\ref{eq:rhos-4} comes from the Fermi level of the 2DEG 
being equal to the bound level of shallow donors in the unionized 
region.

\subsection{Depletion Voltage}
\label{sec:depletion}
In a quantum wire, a 2DEG is usually present at the $z=L$ plane before 
any gate voltage is applied.
When a negative gate voltage is applied to the gates, the density of 
the 2DEG is decreased.
The depletion voltage $-V_{dep}$ is the gate voltage at which electrons 
of the 2DEG are completely depleted from under the gates.
The depletion voltage is an important parameter because it characterizes 
the transition of the system of electrons at the $z=L$ plane from 
two-dimensional to quasi-one-dimensional.

The gate voltage actually measures the energy difference between the 
Fermi level in the gates and the Fermi level in the GaAs at $z=L$.
Therefore, (noting that $E_{z0}=0$ at depletion), the depletion voltage 
is given by
\begin{equation}
eV_{dep} = {e^2 N_{s} \over \varepsilon \varepsilon_{0} }
( l_{c}l_{d} + { l_{d}^2 \over 2} ) - \Phi_{sb}',
\label{eq:Vdep}
\end{equation}
where the first term on the right side gives the total band bending 
in the cap and the doped layers, and $\Phi_{sb}'$ is the Schottky 
barrier of the gate-GaAs contact.
Note that $N_{s}$ should be determined from the measured 2DEG density 
$n_{0}$ according to the ionization regime at the zero gate voltage.
From equation~\ref{eq:Vdep}, the depletion voltage should be 
independent of illumination because $N_{s}$ of the regions in the 
doped layer that are under the gates is not affected by illumination.

\subsection{Pinchoff Voltage}
\label{sec:pinchoff}
The pinchoff voltage $-V_{pinch}$ is the gate voltage at which 
electrons are just completely depleted from the $z=L$ plane in 
the quantum wire.
Therefore, it measures the energy difference between the Fermi level 
of electrons in the gates and the bottom of conduction band at 
the central point $(x=0,z=L)$ of the electron channel.
The calculation of the pinchoff voltage is much more complicated 
than that of the depletion voltage, because it involves the 
electrostatic potential difference between at the point $(x=0,z=L)$ 
and the gates, and depends in an essential way on the fringing 
fields of the capacitors discussed in Section~\ref{sec:regimes}.
The pinchoff voltage is affected by illumination because an 
illumination increases the shallow donor density under the exposed 
semiconductor surface and thus changes the charge distribution.

For the purpose of the calculation below, let the electrostatic 
potential just inside the semiconductor adjacent to the gates 
be zero and $-\varphi (x,z)$ be the potential function (noting
that the system is y-independent).
Then the pinchoff voltage is given by
\begin{equation}
e V_{pinch}= e \varphi (0,L) - \Phi_{sb}',
\label{eq:Vpinch}
\end{equation}
where $e \varphi (0,L)$ is the potential energy at pinchoff of 
an electron at point $(x=0,z=L)$, and 
$\Phi_{sb}'$ is the gate-GaAs contact Schottky barrier.

The calculation of $-\varphi (0,L)$ can be done by using the 
Green's function method with the Dirichlet boundary condition.
The general expression of the potential function for $z \geq 0$
contains two terms that correspond to the contributions from 
the spatial charge and from the boundary, respectively \cite{Jackson}
\begin{eqnarray}
-\varphi(x,z) & = & -\varphi_{1}(x,z) + -\varphi_{2}(x,z) \\
& = & {1 \over 4 \pi \varepsilon \varepsilon_{0} }
\int\!\!\int\!\!\int d^{3}r' \rho ({\bf r'}) G({\bf r},{\bf r'})
- {1 \over 4 \pi} \int\!\!\int_{(z'=0)} \! dx' dy' \varphi({\bf r'}) 
{\partial \over \partial z'} G({\bf r},{\bf r'}), 
\label{eq:varphi}
\end{eqnarray}
where ${\bf r}=(x,y,z)$, ${\bf r'}=(x',y',z')$, $\rho ({\bf r'})$ 
is the spatial charge density, and $G({\bf r},{\bf r'})$ is 
the Green's function, which is given by
\begin{eqnarray}
\nabla'^{2} G({\bf r},{\bf r'}) & = & -4\pi \delta ({\bf r}-{\bf r'}), 
\\ G({\bf r},{\bf r'})\mid_{z'=0} & = & 0.
\end{eqnarray}
Using the image method, the solution of the Green's function is
\begin{eqnarray}
G({\bf r},{\bf r'}) = {1 \over [ (x'-x)^2 + (y'-y)^2 + 
(z'-z)^2 ]^{1/2} } \nonumber \\
- {1 \over [ (x'-x)^2 + (y'-y)^2 + (z'+z)^2 ]^{1/2} }.
\label{eq:G-expression}
\end{eqnarray}

At an arbitrary gate voltage prior to the pinchoff voltage, there 
are electrons present at the $z=L$ plane and there may exist an 
unionized region in the doped layer as shown in FIG.~\ref{fig:regime C}.
Therefore, the spatial charge density $\rho ({\bf r'})$ is not 
known analytically, and $-\varphi(x,z)$ can only be calculated numerically.
At the pinchoff voltage, however, no electrons are present at 
the $z=L$ plane and the shallow donors everywhere in the doped 
layer must be ionized.
(If there were an unionized region, there would have to be 
electrons present at the $z=L$ plane because the bottom of the 
conduction band in the GaAs channel layer is lower than that 
in the doped layer by $E_{off}$.)
Because of this, it is possible to calculate $-\varphi(x,z)$ 
analytically at the pinchoff voltage.

Because the shallow donors are all ionized, before illumination, 
the spatial charge density can be expressed as
\begin{equation}
\rho ({\bf r}) = \left\{ \begin{array}{ll}
eN_{s}, & \mbox{  if $l_{c} \leq z \leq l_{c}+l_{d}$} \\
0, & \mbox{  otherwise}
\end{array} \right. 
\end{equation}
The contribution from the spatial charge, the first term on the 
right side of equation~\ref{eq:varphi}, can thus be calculated 
easily and the result is
\begin{equation}
-\varphi_{1}(x,z) = { eN_{s} \over \varepsilon \varepsilon_{0} }\times
\left\{ \begin{array}{ll}
l_{d} z, & \mbox{ if $z<l_{c}$} \\ 
-{1 \over 2} (z-l_{c}-l_{d})^{2} + l_{c}l_{d} + {1 \over 2}l_{d}^{2}, 
& \mbox{ if $l_{c} \leq z \leq l_{c}+l_{d}$ } \\
l_{c}l_{d} + {1 \over 2}l_{d}^{2}, & \mbox{ if $z>l_{c}+l_{d}$}
\end{array} \right.   
\end{equation}
which is independent of $x$.
For the central point $(x=0,z=L)$,
\begin{equation}
\varphi_{1}(0,L) = {eN_{s} \over \varepsilon \varepsilon_{0} } 
( l_{c}l_{d} + {l_{d}^{2} \over 2} ),
\label{eq:varphi1-1}  
\end{equation}
which gives the total band bending in the cap and the doped layers.
(We have obtained this result previously in Section~\ref{sec:regimes} by 
using consideration of the capacitor.)

After an illumination, the spatial shallow donor density has been 
increased in the doped layer under the exposed surface (Feature 1).
As an approximation, we can take the spatial charge density as 
\begin{equation}
\rho ({\bf r}) = \left\{ \begin{array}{ll}
eN_{sl}, & \mbox{ if $x \leq |w|$ and $l_{c} \leq z \leq l_{c}+l_{d}$} \\
eN_{s}, & \mbox{ if $x>|w|$ and $l_{c} \leq z \leq l_{c}+l_{d}$} \\
0, & \mbox{ otherwise}
\end{array} 
\right.    
\label{eq:Nsl}
\end{equation}
in which $N_{sl}>N_{s}$ because the shallow donor density has 
been increased under the exposed surface. 
$N_{sl}$ can be determined in the same way as $N_{s}$  from the 
2DEG density after illumination at zero gate voltage.

After performing the integration, the potential due to the 
spatial charge after illumination 
can be expressed as
\begin{equation}
\varphi_{1l}(0,L) = \varphi_{1}(0,L) [1-{N_{sl}-N_{s} \over N^{sl}} 
(\alpha_{1}+\alpha_{2})],
\end{equation}
where $\varphi_{1}(0,L)$ is given by equation \ref{eq:varphi1-1}, 
and $\alpha_{1}=L/ \pi w$ and $\alpha_{2}=-L^{3}/ 3 \pi w^{3}$.

Now let us calculate the boundary contribution, the second term 
in equation \ref{eq:varphi}.
For split-gate quantum wires, we have a technical problem with 
potential value at the boundary potential ($z=0$).
Although we know the boundary potential near the gates (which has 
been chosen to be zero here), we do not know exactly how the 
boundary potential is distributed on the exposed surface. 
Strictly speaking, the potential distribution on the exposed 
surface depends on the detailed information of the surface states.
But the physics of surface states is very complicated and a 
calculation including the full details of the surface states is 
not feasible.
However, in studying quantum wires, we find that it is sufficient 
to make some simple assumptions based on the properties of the 
surface states which have been described in Feature~3.

Considering the symmetry of quantum wires, the potential function 
at the exposed surface can be expanded as
\begin{equation}
-\varphi(x,0) = \sum_{k=0}^{\infty}a_{k}x^{2k},\ |x| \leq w
\label{eq:expansion}.
\end{equation}
where $\{ a_{k} \}$ are constant coefficients.
We find it necessary to keep the first two terms in the expansion 
\ref{eq:expansion}.
Such a treatment makes it possible to ensure that the surface
potential is continuous at $x=\pm w$.
This yields
\begin{equation}
-\varphi(x,0) = \left \{ \begin{array}{ll}
V_{0} (1 - x^{2}/w^{2}), & \mbox{ if $|x| \leq w$} \\
0, & \mbox{ if $|x|>w$}
\end{array} 
\right.    
\label{eq:surface potential}
\end{equation}
where $V_{0}$ is a constant and will be determined later.
Substituting equation \ref{eq:surface potential} into the second 
term in equation \ref{eq:varphi}, the boundary contribution is
\begin{equation}
-\varphi_{2}(x,z) = {V_{0} \over \pi w^{2}} [(w^{2}+z^{2}-x^{2}) 
\theta(x,z)
+ xz \ln {(w+x)^{2} + z^{2} \over (w-x)^{2} + z^{2}} -2wz],
\label{eq:phi3-ex2}
\end{equation}
where 
\begin{equation}
\theta(x,z) = \arctan {w-x \over z}+\arctan {w+x \over z},
\end{equation}
which is just the angle that is subtended by the exposed surface 
at the point $(x,z)$.
Therefore,
\begin{equation}
e\varphi_{2} (0,L) ={2e V_{0} \over \pi} [(1 +{L^{2} \over w^{2}}) 
\arctan{w \over L} - {L \over w}].
\label{eq:phi2}
\end{equation}

The boundary contribution $-\varphi_{2}(x,z)$ actually describes 
the potential
that laterally confines the electrons at the $z=L$ plane.
To help visualize this confining potential, we plot 
$e \varphi_{2}(x,z)$ in $z>0$ half space in 
FIG.~\ref{fig:3d potential}.
The intersection of the plot with the $z=L$ plane just gives 
the confining potential well profile.
The larger $L$ is, the more shallow the potential well becomes.

Now $V_{0}$ can be determined by the conservation of 
the total number of surface electrons (Feature~3).
That is
\begin{equation}
\int_{-w}^{w} n^{sur}(x) dx = 2 e w N_{s} l_{1},
\label{eq:conservation}
\end{equation}
in which the right side expresses the linear charge density of 
the exposed surface at zero gate voltage, and the left side
is the linear charge density at the pinchoff voltage.
The area surface density is evaluated based on the calculated
$-\varphi (x,z)$, to yield
\begin{equation}
e n^{sur}(x) = {2 \varepsilon \varepsilon_{0} V_{0} w \over 
\pi w^{2}} [x \ln {(w+x)^{2} \over (w-x)^{2}} - 4 w] - e N_{s} l_{d}.
\label{eq:nsur}
\end{equation} 

Finally, we discuss briefly the relationship between the work 
presented in this section and the earlier work of 
Davies \cite{Davies} who was the first to study the boundary 
contribution to the potential of a quantum wire using the 
Green's function method.
Davies considered only the leading term in the 
expansion~\ref{eq:expansion} of the surface potential.
However, for our purposes this approximation is not 
adequate since it yields a discontinuous potential along 
the surface instead of equation~\ref{eq:surface potential}, 
and as a consequence, a surface charge density for which 
the integral in equation~\ref{eq:conservation} diverges.
By retaining also the second term of the 
expansion~\ref{eq:expansion}, we obtain a continuous surface 
potential and a finite integrated surface charge density~\ref{eq:nsur}. 
This enables us to use the conservation of the surface charge 
at the exposed surface to evaluate the parameter $V_{0}$.

\section{Discussion of a Real Sample}
\label{sec:discussion}
Now let us take a real split-gate quantum wire as an example 
for calculating the depletion and pinchoff voltages using the 
present theory.

The sample quantum wire we consider has the typical structure 
displayed in FIG.~\ref{fig:crossection}.
Grown with MBE on a semi-insulating GaAs substrate, its layers 
in sequence are a 65~nm GaAs buffer, 30 periods of GaAs/AlAs 
superlattice, 900~nm GaAs channel layer, 1.5~nm AlAs and 16~nm 
undoped ${\rm Al_{0.33}Ga_{0.67}As}$ layers as the spacer, 
40~nm Si-doped ${\rm Al_{0.33}Ga_{0.67}As}$ layer with donor 
concentration of ${\rm1.1 \times 10^{18}\ cm^{-3}}$, and 
18~nm GaAs cap layer with normal surface (100).
On top of the GaAs cap, two separated gate bars of titanium 
are applied using electron beam lithography.
The gate bars have a spatial separation of 200~nm and width 
of 200~nm.

Analysis after growth shows that the undoped 
${\rm Al_{0.33}Ga_{0.67}As}$ layer of the spacer is 14.5~nm 
instead of the expected value 16~nm.
This suggests that all of the actual thicknesses should be 
reduced by 10\% from their expected values.
Correspondingly, the concentration of the Si donors in 
the doped ${\rm Al_{0.33}Ga_{0.67}As}$ layer should be 
increased by 10\% so as to keep the nominal total number of donors.
The parameter values that we use in our calculations are listed in
TABLE~\ref{tab:parameters}.

According to equations \ref{eq:Nalpha}, \ref{eq:Nbeta-1}, 
\ref{eq:Nbeta-2}, \ref{eq:Nbeta-3}, and \ref{eq:nbeta}, 
the calculated critical values that separate the ionization 
regimes of this sample are 
$N_{\alpha}=0.45 {\rm \times10^{18}\ cm^{-3}}$, 
$N_{\beta}=0.80 {\rm \times10^{18}\ cm^{-3}}$, 
and $n_{\beta}=6.03 {\rm \times 10^{11}\ cm^{-2}}$.

Now let us consider three situations of the quantum wire: 
before illumination, after one illumination, and after 
many illuminations (i.e. after saturation with a red 
light emitting diode).
Corresponding to these three situations, the measured 
densities of the 2DEG at zero gate voltage are 
${\rm 3.40 \times10^{11}\ cm^{-2}}$, 
${\rm 5.49 \times10^{11}\ cm^{-2}}$, 
and ${\rm 6.25 \times10^{11}\ cm^{-2}}$, respectively.
Comparing these measured values to the calculated 
critical value $n_{\beta}=6.03 {\rm \times 10^{11}\ cm^{-2}}$ 
and referring to TABLE~\ref{tab:parameters}, 
the ionization regimes of this quantum wire 
before illumination, after one illumination, 
and after many illuminations are in B, B, and C, respectively.
The corresponding shallow donor densities and the depletion 
voltage and pinchoff voltages can therefore be calculated 
based on the formalism in Section~\ref{sec:model}.
The calculated results are presented in TABLE~\ref{tab:results}.

Experimentally, the depletion and pinchoff voltages can be known 
from the measured longitudinal (y-direction) resistances against 
the gate voltage.\ \cite{van_Wees,Wharam}
The measured resistance curves of the sample quantum wire 
are displayed in FIG.~\ref{fig:resistances}.
Curves a, b, and c correspond to the resistances varying with 
the gate voltage before illumination, after one illumination, 
and after many illuminations, respectively.
The depletion voltages for curves a, b, and c are $-$0.33 V, 
$-$0.35 V and $-$0.37 V, respectively, which agree very well
with the calculated results $-$0.33 V.
(The negative increases of depletion voltage upon illumination 
can be explained by the fact that the gate bars are very 
narrow and therefore some illuminating photons may penetrate 
into the regions under the gates and excite the deep donor there.)
The pinchoff voltages read from curves a, b, and c are about $-$0.55 V, 
$-$0.86 V, and $-$1.33 V, respectively, which are fairly close to
their corresponding calculated results $-$0.53 V, 
$-$0.80 V, and $-$1.43 V.

In conclusion, this paper presents an electrostatic model of 
split-gate quantum wires and sets up a general formalism 
that is applicable  both before and after illumination.
For any split-gate quantum wire, given its geometric parameters 
and its measured 2DEG density at zero gate voltage, additional 
information such as its ionization state, shallow donor density, 
depletion voltage, and pinchoff voltage can be calculated based 
on the model.
While contributing to our understanding of the electrostatic 
characteristics of quantum wires, this model suggests a potential
approach for studying the electrodynamic and time-dependent 
processes in quantum wires.
The theory of this paper should also provide a tool for study 
other gated nanostructures such as multiple constrictions 
\cite{Smith,Hwang,Schmit,Simpson} and quantum dots
\cite{Reed,Kouvenhoven,Lorke}.

We would like to acknowledge helpful discussions with C. J. B. Ford,
MBE material grown by 
P. T. Coleridge and fabrication assistance from P. Chow-Chong,
M. Davies, P. Marshall, R. P. Taylor, and R. Barber.
This work was supported by the Natural Sciences and Engineering 
Research Council of Canada and the Centre for Systems Science at 
Simon Fraser University.

\figure{Crossection of a typical split-gate quantum wire and the 
coordinate frame chosen for
calculations.
\label{fig:crossection}}

\figure{A possible band bending structure within quantum wires.
The curve shows the bottom of the conduction band along the z-axis.
$E_{off}$ is the band offset at the interfaces between GaAs and ${\rm
Al_{x}Ga_{1-x}As}$.
\label{fig:regime C}}

\figure{Band bending structures for ionization regime A (a) and 
ionization regime B (b), see text.
\label{fig:regimes AB}}

\figure{A Three-dimensional representation of the potential energy 
$e \varphi_{2}(x,z)$ due to the
boundary contribution for $z \geq 0$.
Used parameters are $w=100$ (nm) and $V_{0}=1$ (arbitrary unit).
\label{fig:3d potential}}

\figure{The measured resistances of the sample quantum wire.
Curves a, b, and c are for the cases of the wire before 
illumination, after one illumination, and after many illuminations, 
respectively.
\label{fig:resistances}}

\begin{table}
\caption{Criteria for identifying different ionization regimes of 
the doped layer under the exposed surface in quantum wires at 
zero gate voltage.}
\begin{tabular}{ccc}
Ionization regime & By shallow donor density $N_{s}$ & By 2DEG density
$n_{0}$ \\ 
\tableline
A & $N_{s} < N_{\alpha}$ & $n_{0} = 0$ \\
B & $N_{\alpha} < N_{s} < N_{\beta}$  & $0<n_{0} < n_{\beta}$ \\
C & $N_{s} > N_{\beta}$ & $n_{0} > n_{\beta}$ \\
\end{tabular}
\label{tab:ionizations}
\end{table}

\begin{table} \caption{Parameters used in
calculations for the real sample of quantum wire} 
\begin{tabular}{cccc}
Description & Notation & Value & Unit \\ 
\tableline
gate separation & $2w$ & 200 & nm \\
GaAs cap layer & $l_{c}$ &16.2 & nm \\
doped ${\rm Al_{0.33}Ga_{0.67}As}$ layer & $l_{d}$ & 36 & nm \\
${\rm Al_{0.33}Ga_{0.67}As}$ and AlAs spacer & $l_{s}$ & 15.75 & nm \\
effective mass & $m^{*}$ & 0.067 & $m_{e}$ \\
dielectric constant of GaAs & $\varepsilon$ & 12.5 & \\
Schottky barrier of GaAs surface (100) \cite{Chiang}& 
$\Phi_{sb}$ & 0.80 & eV \\
Schottky barrier of Ti-GaAs contact \cite{Waldrop} & 
$\Phi_{sb}'$ & 0.83 & eV \\
band offset \cite{Okumura} & $E_{off}$ & 0.2 & eV \\
z-direction energy interval \cite{Harris} & $E_{z0}$ & 0.04 & eV \\
\end{tabular}
\label{tab:parameters}
\end{table}

\begin{table}
\caption{Calculated results of the real quantum wire}
\begin{tabular}{ccccc}
Parameter & Before ill. & After one ill. & After many ill. & Unit \\
\tableline 
Ionization regime & B & B & C & \\ 
measured 2DEG density $n_{0}$ & 3.40 & 5.49 & 6.25 & 
${\rm 10^{11}\ cm^{-2}}$ \\
shallow donor density & 0.65 & 0.77 & 1.02 & 
${\rm 10^{18}\ cm^{-3}}$ \\ 
$l_{1}$ & 30.89 & 28.87 & 24.03  & nm \\
$l_{2}$ & 5.11 & 7.13 & 6.11  & nm \\
$l_{3}$ & 0 & 0 & 5.86  & nm \\
calculated $-V_{dep}$ & $-$0.33 & $-$0.33 & $-$0.33 & V \\
calculated $-V_{pinch}$ & $-$0.53 & $-$0.80 & $-$1.43 & V \\
\end{tabular} 
\label{tab:results}
\end{table}


\begin{references}
\bibitem{2DEG} Experimentally, the bulk density of the 2DEG can be 
determined by Shubnikov de Haas measurements.
\bibitem{expansion} Keeping more terms in the expansion will 
modify the calculated distribution of the surface charge density.
However, such a change (for fixed total surface charge) has no 
significant effect on the potential at the $z=L$ plane because 
this plane is far away from the surface.
\end{references}
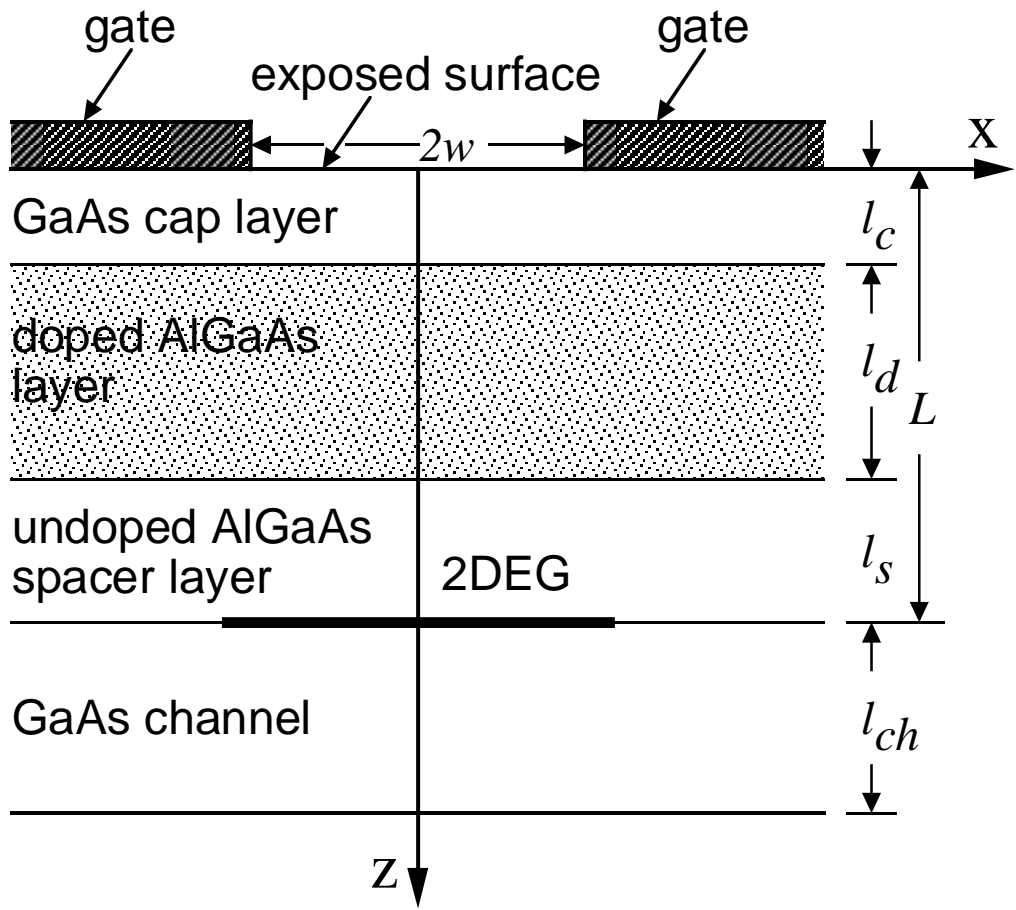
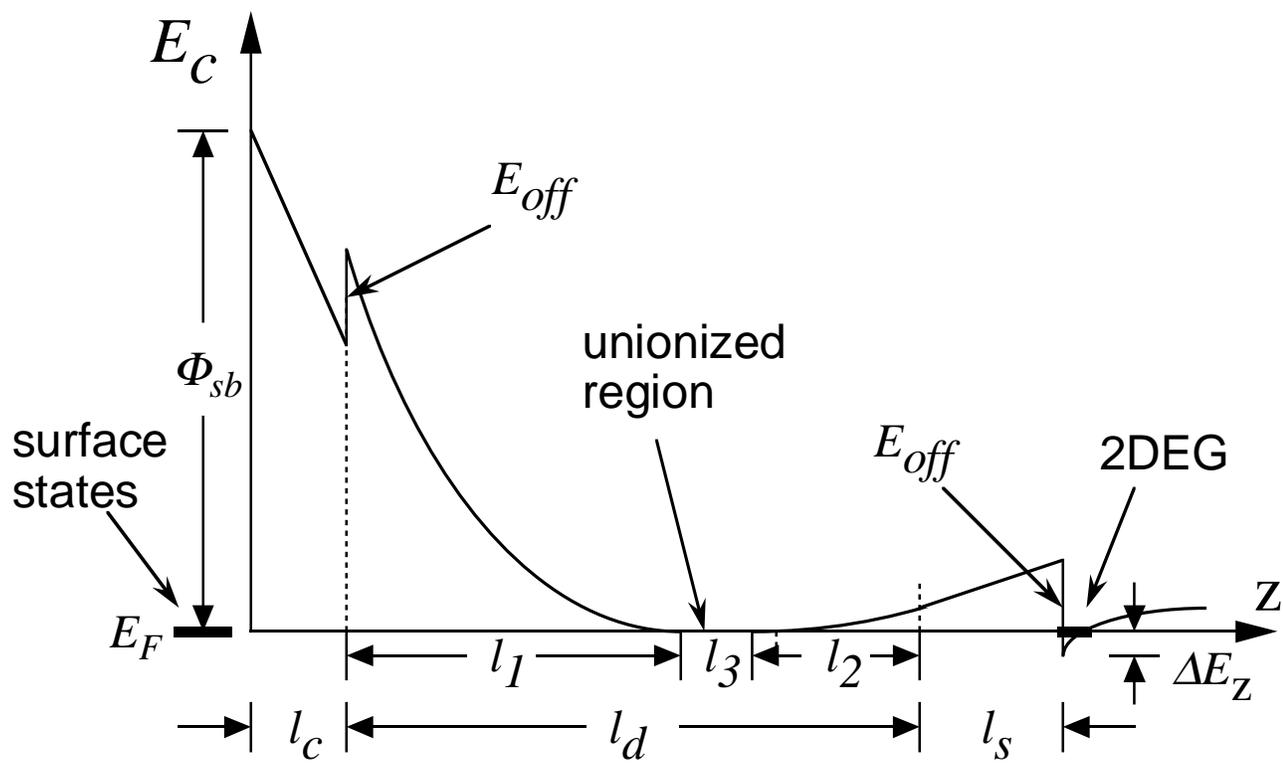
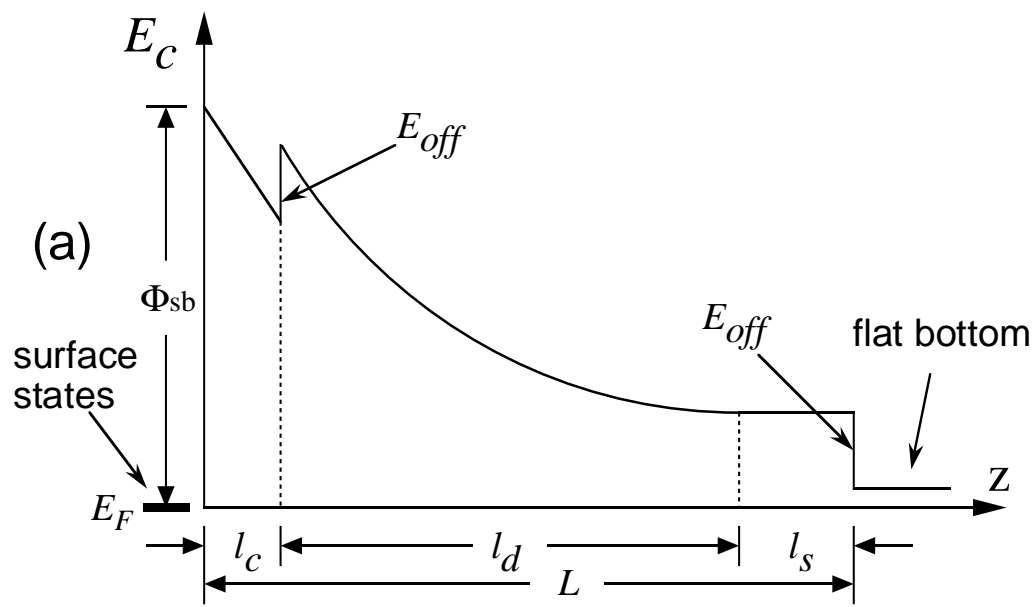
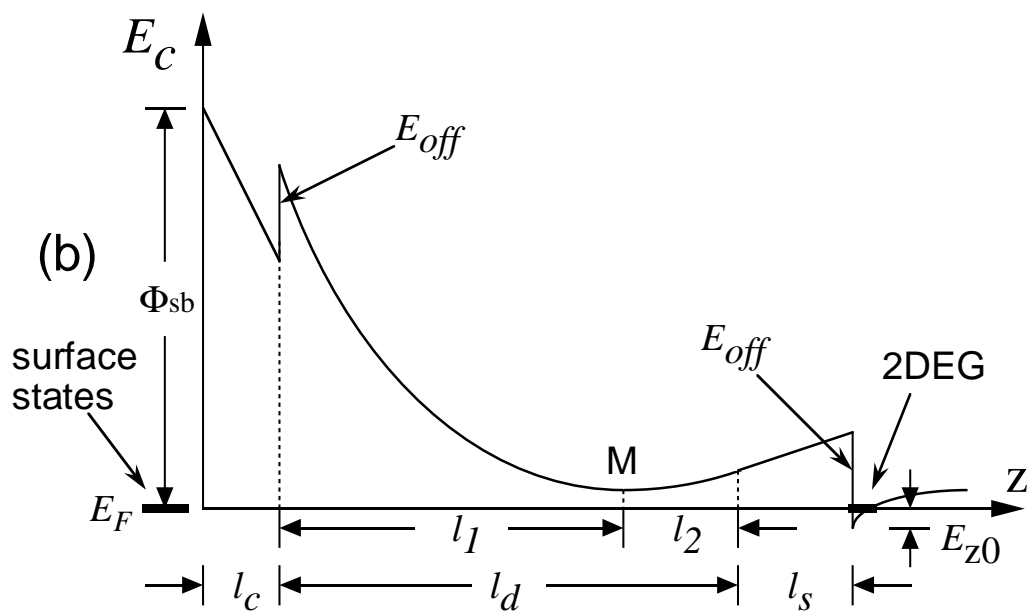
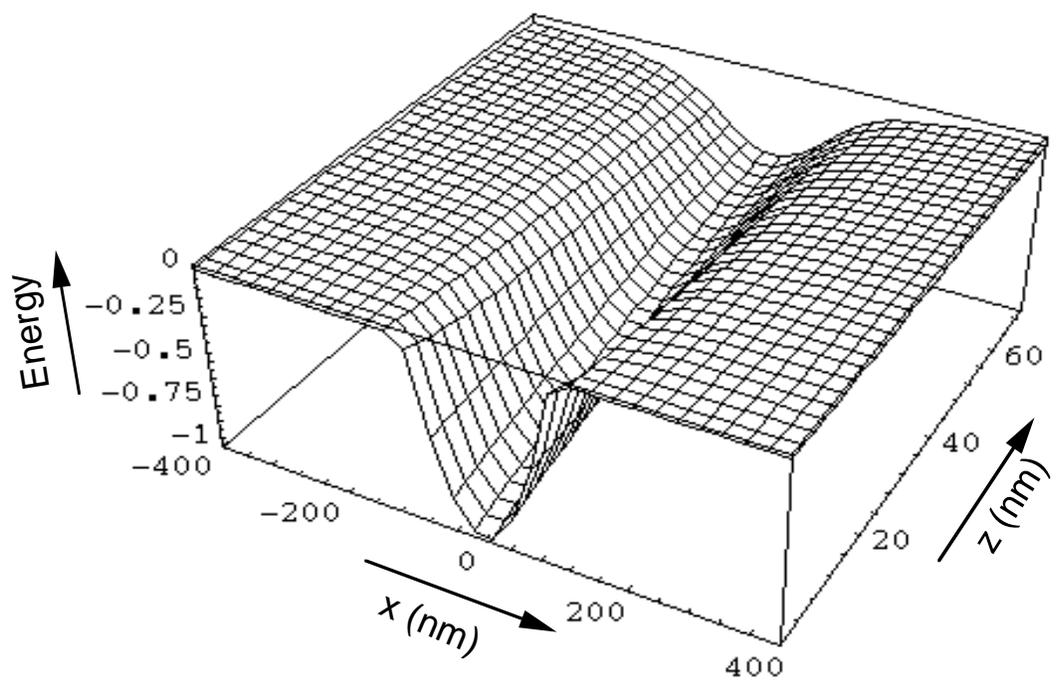
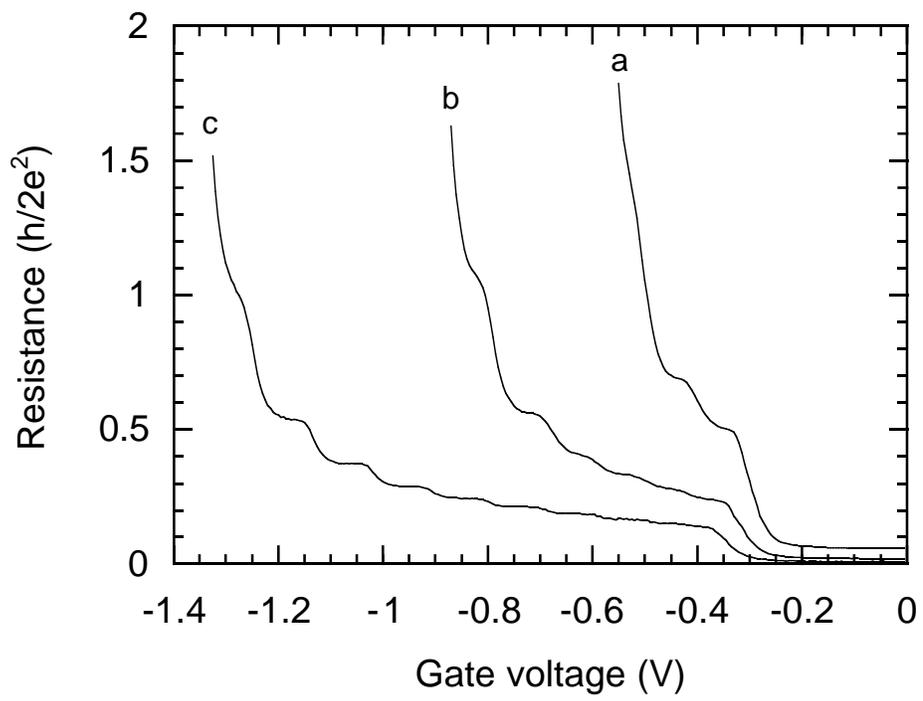
\end{document}